# Using Social Network and Semantic Analysis to Analyze Online Travel Forums and Forecast Tourism Demand

Fronzetti Colladon, A., Guardabascio, B., & Innarella, R.

Fronzetti Colladon, A., Guardabascio, B., & Innarella, R. (2019). Using Social Network and Semantic Analysis to Analyze Online Travel Forums and Forecast Tourism Demand. Decision Support Systems, 123, August 2019, 113075. https://doi.org/10.1016/j.dss.2019.113075

# Using Social Network and Semantic Analysis to Analyze Online Travel Forums and Forecast Tourism Demand

Fronzetti Colladon, A., Guardabascio, B., & Innarella, R

**Abstract**

Forecasting tourism demand has important implications for both policy makers and companies operating in the tourism industry. In this research, we applied methods and tools of social network and semantic analysis to study user-generated content retrieved from online communities which interacted on the TripAdvisor travel forum. We analyzed the forums of 7 major European capital cities, over a period of 10 years, collecting more than 2,660,000 posts, written by about 147,000 users. We present a new methodology of analysis of tourism-related big data and a set of variables which could be integrated into traditional forecasting models. We implemented Factor Augmented Autoregressive and Bridge models with social network and semantic variables which often lead to a better forecasting performance than univariate models and models based on Google Trend data. Forum language complexity and the centralization of the communication network – i.e. the presence of eminent contributors – were the variables that contributed more to the forecasting of international airport arrivals.

## 1. Introduction

The tourism industry represents an extremely complex business scenario, where companies carrying out very different activities integrate their products and services – these comprise travel agencies, tour operators, restaurants, hotels, transportations providers, etc... Products and services can be sold both individually or in holiday packages [1]. Accessing local knowledge is a fundamental step when people are planning a trip. This information can be provided by travel agencies, personal acquaintances, guide books, or by the web. With the rapid evolution of the internet and connected devices, such as laptops and mobile phones, the information that people can access on the web has dramatically increased [2], also producing a revolution in the tourism industry. New technologies and online services changed the way tourists relate with travel agents and the way they organize new trips [3]: for example, people can now easily use the web to look for the cheapest flights, compare thousands of hotels, book their access to a museum, or reserve a table at a restaurant. Consequently, the numbers of clients in the industry increased as well as the amount of information they can access [4,5]. Moreover, operators can now offer their products and services without intermediaries, thus having the possibility to reduce the final price. Competition is always stronger and marketing strategies can leverage on a better knowledge of the consumer to offer personalized products [6]. Companies can now increase their profits through insights coming from the analysis of search queries on Google, or of the content of online reviews [7]. The consumer is now even smarter and more aware of the tricks behind some marketing campaigns. Therefore, many people prefer to rely on the judgement provided by their peers, more than on the information they find on companies' websites. The online interaction on social networks, or on dedicated platforms, makes people feel part of a group [8]; many of them get a sense of reward when they can share their knowledge and help others [9]. Accordingly, online reviews and user-generated content acquired a great importance and made the success of very well-known websites like TripAdvisor, also confirming their usefulness while making tourism demand predictions [10–12]. Big data shared on online social networks can help anticipate rapid changes in tourist preferences and popularity trends of destinations and local attractions; this can be achieved by both analyzing the topics emerging from the online discourse and studying the interaction dynamics among users [13–17].

Following this trend, we propose the analysis of the online travel forums included in one of the world's leading tourism platforms, TripAdvisor, by using methods and tools from social network and semantic analysis [18,19]. The objective is to discuss the usefulness of variables extracted from the study of online communities, in order to forecast international arrivals in the airports of European capital cities. Our contribution is based on the investigation of both the content of people's posts and their social interactions, with the idea that a more active online community, where knowledge-sharing is supported by functional social dynamics, can be predictive of a higher number of arrivals. We present new variables that are relatively easy to extract and monitor from online sources and which could be integrated in other existing forecasting models to improve their accuracy. In this study, we test our methodology considering the last 10 years of the online discourse on TripAdvisor's forums, focusing our attention on 7 major European capital cities. To be consistent with the analysis of the language use, we limited our sample to posts written in English. Nonetheless, future research could replicate our methodology considering other online sources, different languages, and focusing on other predictions (such as the number of visitors to museums or other specific tourist attractions). It is important to consider that our study is exploratory for a part. We prove the informative value of semantic and social network indicators, without the ambition of providing full explanation about the reasons behind their influence on the forecasts made for each city. This would require a new dedicated research, which we advocate for the future.

Forecasting tourism demand has significant policy implications; insights from our analysis are useful both for decision makers at a regional and country level and for companies operating in the tourism industry [20,21]. Better predictions can help local companies and policy makers to allocate resources, define pricing policies and implement business plans. More accurate predictions reduce the risk of misplanning, and can be vital for the growth of tourism-dependent economies, both at a local and at a national level [22,23]. Our study also contributes to the literature about tourism forecasting, presenting a new methodological approach and new metrics – based on the social network and semantic analysis of big data – which go beyond the study of online reviews or web search activity [22,24].

## 2. Forecasting Tourism Demand

Big data and the development of information and communication technologies have a great importance for the tourism industry, as internet is a preferred knowledge source for tourists and one of the most important drivers of tourism demand [4,25,26]. Accordingly, new buzzwords are emerging, such as 'smart tourism' – a concept used to "describe the increasing reliance of tourism destinations, their industries and their tourists on emerging forms of ICT that allow for massive amounts of data to be transformed into value propositions" [27]. New data can now be acquired analyzing tourist interactions on social media websites or their use of mobile applications which enhance their travel experience [28–30]. Big data analytics can provide new knowledge about destination choices [31], support strategic decision-making in tourism destination management [32], and help the forecasting of new arrivals [33,34]. In this context, social media and online reviews play a significant role, as they support information search, decision-making and knowledge exchange for tourists [34]. For the companies operating in the tourism industry, social media represent a means to communicate with customers and a place for the implementation of a good part of the marketing strategy [35]. Online travel forums are used by tourists who have specific questions, which are not usually answered in common reviews of tourist attractions: forums reveal specific information needs and their link with prospective destinations [36].

In this study the authors follow a big data approach to extract information from the TripAdvisor travel forum, and measure new variables which could help in forecasting tourist arrivals. Forecasting tourism demand has been a major topic of research in the past decades [37–39]; scholars used a wide range of techniques, with no single model succeeding in outperforming the others in all situations [20]. Some studies focused their attention on the effects that new communication channels, especially social media, have on tourist decisions and choice of destinations [10,40] – for example, Saparks and Browning [24] studied the impact of online reviews on hotel bookings; others researchers investigated the information needs that bring people to generate questions on online travel forums [36].

Tourism demand can be measured using different proxies, such as the number of nights spent in accommodation establishments or the number of visa requirements. Many studies focused on tourists arrivals and provided predictions based on time series and seasonal trends [41,42]. Considering online sources to help these predictions is not new. Some scholars inferred tourism demand from an analysis of search engine and web traffic data [43,44]. Li, Pan, Law and Huang [45] developed a composite search index to more efficiently analyze search query volumes and improve the forecasting accuracy of Chinese tourism demand. Similarly, Yang, Pan, Evans and Lv [33], used autoregressive models combined with search query data. Artola, Pinto and de Pedraza García [46] proved that traditional models can be improved by using data from Google Trends. Choi and Varian [47] carried out a very similar research, using again Google Trends to predict visitors to Hong Kong. Also Bangwayo-Skeete and Skeete [22] supported the idea that Google Trends can help outperform conventional time series models. Gunter and Önder [48], instead, used Google Analytics to predict city arrivals in Vienna.

Recent works proposed methods which combine different data sources and techniques to improve models accuracy [49]. Sun et al. [50], for example, combined data mining and models based on Markov chains. Other scholars examined big data, combining multiple online sources – such as price levels and web traffic – to make predictions [51]. We agree with the importance of using combined approaches [52] and data sources – and maintain the need of finding new variables which can be integrated in existing models; these variables should be also reasonably easy to extract quite in real time.

### 2.1. Exploring online community dynamics to predict tourism demand

Fewer studies used online travel forums data to make predictions. Dali and Yutaka [53], for example, looked at the most recurring words in a Chinese forum, to forecast Chinese people traveling to Japan. To the extent of our knowledge, there are also few studies dealing with social network analysis and prediction of tourism demand. Indeed, the use of social network analysis in tourism is still scarce and recent [54]. With this research, we try to fill this gap. We discuss the role of social network and semantic variables that can be extracted from online big data sources – in our case, the TripAdvisor travel forum – to support the forecasting of international airport arrivals.

We chose to analyze online forums instead of TripAdvisor's reviews, for two main reasons: firstly, to study the discourse about European capital cities overall, without limiting our attention to single tourist services or attractions; secondly, because the effects of reviews on tourist behavior has already been explored by many scholars [55–58]. Indeed, the study of online reviews has sometimes to face the problem of deceptive content, generated by people who share false experiences and judgements to promote local business [59].

The success of an online community depends on many factors such as its level of activity, the presence of rotating leaders and the speed at which users get answers to their questions [19,60]. A community with many active members and posts, where more answers are given to people's questions, is usually more popular than a less participated online group. Koh and Kim [61] proved that knowledge-sharing activity predicts both community participation and promotion. In addition, if the online content is accessible without a registration, this leads to a better indexing on search engines thus attracting more members [62]. Knowledge sharing activities can also be supported by the presence of informal moderators, who keep different social groups together and offer eminent contributions to the discourse [63]. In general, when the users' level of expertise is higher, one could expect more rapid and effective answers to people's questions [64]. In terms of social network structure, the presence of eminent contributors usually translates into higher network centralization [60,65]. In terms of rotating leadership and democratic participation to the community life the picture is still open to debate: on one hand, Antonacci et al. [60] proved the importance of rotating leaders to support participation and growth of virtual communities of practice; on the other hand, Gloor et al. [66] showed that, in more operational contexts, the presence of steady leaders – who keep static positions and use a simple language – is appreciated by knowledge-seeking clients. The use of language is another dimension worth to be explored, not only with regard to complexity. Yin, Bond and Zhang [67] showed that the analysis of positive and negative emotions embedded in review texts can be far more informative than ratings. Salehan and Kim [68] showed that online reviews with a neutral sentiment are perceived as more useful. Accordingly, we expect that forum posts with overly positive sentiment could be perceived as suspicious and less informative by perspective tourists.

Given the influence that online travel communities can have on choices of prospective tourists [53], it is important to understand and measure their dynamics, to see if information can be extracted to make meaningful forecasts. In this study, we use the framework proposed by Gloor and colleagues [19,60,66] which suggests considering three dimensions for a comprehensive analysis of online social interactions: degree of interactivity, degree of connectivity and language use. This implies using methods and tools of Social Network and Semantic Analysis to investigate: the social structure of interaction, i.e. the shape of relationship among community members and, for example, the presence of central leaders; the evolution of this structure over time and metrics of interactivity, such as the average response time to received messages; the style of the language used in online conversations measuring, for example, its positivity or complexity.

Compared to the research about online reviews, the study of online communities to forecast tourism demand is relatively new. As a consequence, we carried out an explorative analysis to discover the most significant variables which could be used to forecast international airport arrivals.

**3. Methodology**

We looked for online data which could be relatively easy and fast to crawl and which could be helpful in predicting the number of visitors to touristic destinations in Europe. Specifically, we focused our experiment on the forecasting of international visitors to seven European capitals, analyzing the online forums of the TripAdvisor website. We chose TripAdvisor as this is the leading tourism online platform, active since February 2000 and used all over the world. In 2017 it counted 535 million users and included reviews and information about 7.3 million restaurants, accommodations, airlines and tourism attractions[1]. The website, available in multiple languages, counts more than 455 million unique visitors every month and has the power to significantly drive and influence tourist decisions. This platform includes an online forum (also accessible to non-registered users) where people can interact by exchanging travel tips and opinions and by sharing personal experiences. This forum deals with topics tightly connected to our research question, it is rich in information and user interaction, and has a high number of posts: as a result, it is a suitable candidate for our analysis [69].

In order to extract forum data, we developed a specific web crawler using the Java programming language. The crawler was able to parse html pages and extract information of interest, with associated timestamps to allow a longitudinal analysis. We conducted our experiment analyzing more than 2,660,000 forum posts, written by more than 147,000 users, considering a time period of ten years (from January 2007 to December 2016). We did not collect antecedent posts as the first forum interactions were in September 2004 and we wanted to be sure to skip the forum startup phase. Our analysis was restricted to posts written using the English language for two main reasons: firstly, to be consistent in the measurement of semantic variables; secondly, because English was the most used language for the exchange of opinions among tourists of different nationalities. In addition to forum interactions, we analyzed profile pages where information about participants – such as their gender, age and number of posts/reviews – were available.

---

[1] https://tripadvisor.mediaroom.com/us-about-us

For the selection of the seven European capitals, we considered the top European nations according to the EUROSTAT[2] ranking on the number of nights spent in tourist accommodation establishments for the year 2016. Subsequently, we selected those capital cities for which we found a significant number of forum posts on TripAdvisor in the past ten years (more than 100,000 posts overall, at least 10,000 per year). Cities selected with this procedure would have been the same if considering the European capital cities with the highest number of international airport arrivals[3]. Due to data quality issues, we could not analyze three cities we originally selected: Athens, London and Rome. For these cities the crawler produced a significant amount of incomplete or inconsistent data – as the website API returned errors or because the html structure of the webpages resulted inconsistent (or changed) during the collection process. Therefore, to avoid introducing biases in the analysis, we preferred working on a sample of 7 cities, for which we could collect verified data of good quality. The capitals included in the study were: Amsterdam, Berlin, Lisbon, Madrid, Paris, Prague and Vienna. We analyzed 7 separate datasets, as each city had a dedicated travel forum on the online platform, organized in forum topics. Users could either open new topics or comment on existing ones. Table 1 shows the total number of forum posts and users for each city, as extracted from the crawler. We see that Paris had the highest participation.

| City | Number of Posts | Number of Users |
|---|---|---|
| Amsterdam | 120,055 | 13,020 |
| Berlin | 156,452 | 12,892 |
| Lisbon | 103,405 | 10,414 |
| Madrid | 189,760 | 14,629 |
| Paris | 1,670,754 | 67,084 |
| Prague | 280,461 | 17,922 |
| Vienna | 146,414 | 11,143 |

**Table 1.** Number of forum posts

We here present the list of variables we could measure and include in the study. The measurement of each variable was repeated on a monthly basis, for each capital city.

*Percentage Male*. It is the proportion of male users who posted in the forum.

---

[2] http://ec.europa.eu/eurostat/documents/2995521/7822893/4-24012017-AP-EN.pdf/922150f7-b642-418d-ab42-9867347d5439s

[3] http://ec.europa.eu/eurostat/web/transport/data/database

*Average Age*. It is the average age of users who posted in the forum.

*Users Level*. Each user activity on TripAdvisor is rewarded by a specific number of points – for example, users get 100 points for writing a review, 30 points for uploading a photo and 20 points for writing a forum post. Points translate into levels (ranging from 0 to 6, where level 1 is obtained at 300 points and level 6 at 10,000 points or more). Users who largely contribute to the website are awarded with a higher level, which reflects their reputation and partially their expertise. Users level is calculated as the sum of individual levels, considering those users interacting in a city forum.

*Users Photos*. It is the sum of the total number of photos uploaded on TripAdvisor by the users who were active in a city forum.

As a proxy for the number of international tourists traveling to a capital city, we considered the number of international arrivals in that city airport (excluding transit passengers), as extracted from the EUROSTAT[4] database. Even if considering airport arrivals has been done in previous studies [70] and air transport and tourism proved to be interlinked [71], our choice can have some potential limitations as people could be traveling for work and not for tourism-related reasons. Moreover, tourists could access a capital city by other means of transport. Some of these limitations are common to other possible proxies for the level of tourism – for example, if the number of nights spent in tourist accommodation establishments are taken into account, there would be the problem of including people staying in hotels for work purposes. Moreover, the number of nights spent in a city does not necessarily reflect the number of people who visited that city, due to the variability of the time spent in the city by each tourist [70]. Another indicator – which has been used in the past [72] – is the number of visa requirements, which is however very difficult to associate to the number of visitors to a specific city and is therefore more appropriate when carrying out an analysis at a country level. In addition, European tourists often do not need a visa to access other countries in Europe. Accordingly, we maintain that our choice of selecting international airport arrivals as the dependent variable of our study is not completely free from possible biases, but it can still represent a good proxy of tourism demand.

---

[4] http://ec.europa.eu/eurostat/web/transport/data/database

This choice is consistent with other studies [70] which already proved that level of tourism is associated to airport arrivals [73].

### 3.1. Social Network Data

Collecting forum data was important as it allowed to map the interaction dynamics within the online communities. Thanks to our crawler we were able to extract the social network of each city forum, where users are nodes, connected by arcs which represent their interactions (answers): if user A answers to a post of user B, there is an arc starting at node A and terminating at node B. The typical behavior was to open a new thread for each new question. Therefore, answers in a thread were mainly related to the original post. User A and B represent one of the many network dyads. The single user could either open new discussion threads or add comments to threads created by others. Multiple replies were possible and the same user could reply several times the same post. The final network has been obtained considering all interactions among users.

Figure 1 shows an example of social network for each city in October 2016 (visualizing the network for the ten years was computationally not viable, given its very big size). Network size is consistent with the rankings reported in *Table 1* – with Paris having the most participated forum.

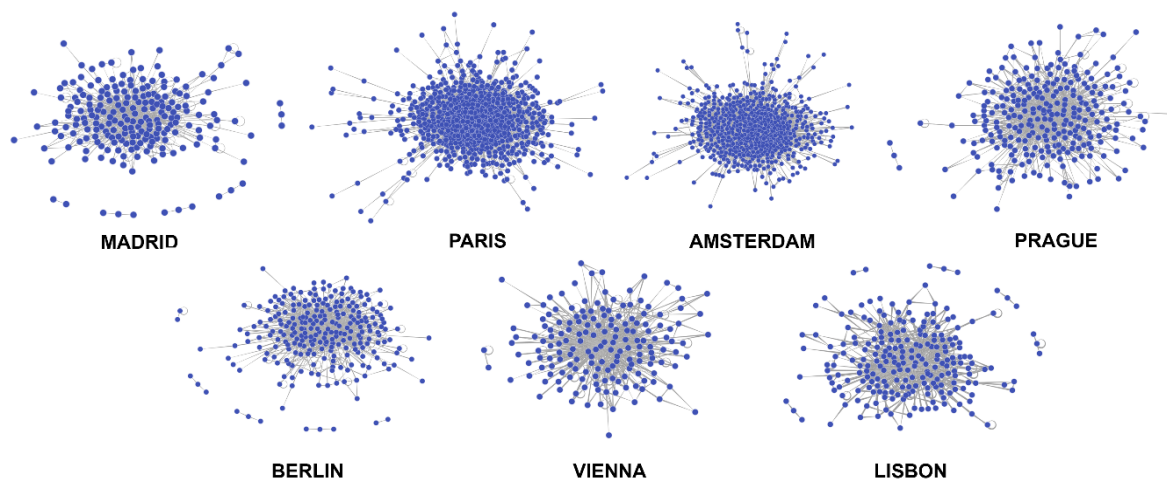

**Figure 1**. Social Network of Online Communities in October 2016.

The contribution offered by this research is based on the exploration of online social interaction in travel forums to identify variables which can help forecasting tourist arrivals. Specifically, we investigated social dynamics according to the framework proposed by Gloor and colleagues [66], which is based on the measurement of the degree of connectivity and interactivity in online communities and on the analysis of language use.

Social structure (connectivity) was studied considering the two well-known metrics of *Group Degree Centrality* and *Group Betweenness Centrality* [18]. Degree centrality is a measure of the number of direct connections of each user; it answers to the question: "how many other users did he/she interacted directly with?". When measured at the group level, it shows how much variation there is in degree centrality scores of individuals. If a network is dominated by a central actor, connected to all others who do not share connections among them, the group degree centrality is maximum and equal to 1 [18]. Betweenness centrality, on the other hand, is a measure that goes beyond direct links and shows how frequently a node lies in the paths that interconnect the other nodes; this measure can often be considered as a proxy of the amount of information that passes through a specific social actor [18,74]. Similarly to group degree centrality, group betweennes centrality expresses the heterogeneity of betweenness centrality scores, and it reaches the maximum value of 1, if the network is close to a star graph, where a central actor interconnects all his/her peers [18].

Interactivity has been studied by considering the number of new users, the levels of activity and the *Average Response Time (ART)* taken by users to answer comments or questions (measured in hours). *Activity* counts the number of network links generated by the users' posts. The *New Users* variable counts the number of new users joining an online city forum.

In addition, we calculated a group level metric which expresses the *Rotating Leadership* of community members, operationalized as the count of their oscillations in betweenees centrality [75]. A community where members occupy static positions - for example for the presence of eminent contributors who share their unique knowledge - has zero or few oscillations; on the other hand, when community members support the active participation and involvement of other users, they rotate more, sharing their leadership and making the interactions more 'democratic'.

The use of language was studied along the dimensions of language *Sentiment* and *Complexity*. Sentiment is a measure expressing the positivity or negativity of community posts; it ranges from 0 to 1, where 0 represents very negative posts and 1 very positive ones. The calculation was made by using the machine learning algorithm included in the software Condor [19]; we used the same software to calculate also language complexity, based on the likelihood distribution of words within a post, as illustrated in the work of Brönnimann [76]. Briefly, complexity is the probability of each word to appear in the text based on the term frequency/inverse document frequency (TF-IDF) information retrieval metric:

$$Complexity = \frac{1}{n} \sum_{w \in V} q(w) \log \frac{1}{p(w)}$$

where *n* is the total number of words within a post, *V* is the vocabulary of words that appear in the post, *q(w)* is the frequency of word *w*, *p(w)* is the probability of word *w* to appear in a post, and *log 1/p(w)* is the inverse document frequency of word *w* in the corpus.

Lastly, in order to compare the outcomes of our model with past research, we collected two additional variables named *Google Trend Flights* and *Google Trend Holidays;* these variables correspond to the Google Trend search volume index for the search queries made by the name of a city followed (or preceded) by the word "flights" or the word "holidays" respectively. This choice is consistent with previous studies [77–79], and detailed in the work of Artola et al. [46] who also examined the limitations of this choice. Artola and colleagues showed that using this variables can significantly improve the prediction of tourism inflows. Here we do not dwell on these variables and findings, but use them for comparative purposes.

Table 2 summarizes the variables which we used to forecast international arrivals.

| Variable Name | Brief Description |
|---|---|
| Users Photos | Sum of the total number of photos uploaded on TripAdvisor by the users who were active in a city forum. |
| Users Level | Level attributed to each user on TripAdvisor (summed for each city forum). Depends on user experience (number of reviews, forum posts, uploaded photos). Users who largely contribute to the website are awarded with a higher level, which reflects their reputation and partially their expertise. |
| Percentage Male | Percentage of male users in a forum. |
| Average Age | Average age of users who posted in the forum. |
| Activity | Number of social network links in a forum (generated by comments/answers to users' posts). |
| Group Betweenness Centrality | Expresses the heterogeneity in betweenness centrality scores, which are a proxy of the brokerage power of users: they show how frequently a user is in-between the network paths that interconnect her/his peers [18,80]. |
| Group Degree Centrality | Measures how much variation there is in degree centrality scores of users, i.e. in their number of direct connections (the number of different people they interact with) [18,80]. |
| Rotating Leadership | Sum of users' oscillations in betweenness centrality [75]. A community where interaction is more 'democratic' - as members occupy less static positions - has more oscillations, which is usually beneficial to its participation and growth [60]. |
| Sentiment | Measures the positivity or negativity of the language used, with values in the range [0,1]. Neutral posts have a score of 0.5; higher scores indicate a more positive language [19,81]. |
| Complexity | Measures the complexity of the language used, with more complex posts having a higher score [76]. |
| Average Response Time | Average time taken by users to answer comments or questions (measured in hours). |
| New Users | Counts the number of new users joining a forum. |
| Google Trend Flights | Google Trend search volume index, for the search queries made by the name of a city followed (or preceded) by the word "flights". |
| Google Trend Holidays | Google Trend search volume index, for the search queries made by the name of a city followed (or preceded) by the word "holidays". |

**Table 2.** Variables used to forecast international arrivals.

### 3.2. Forecasting Model

Let us suppose that the scalar time series to forecast $y_t$, is generated by the following autoregressive model (AR):

$$y_{t+h} = \sum_{i=1}^{p} \varphi_i y_{t+1-i} + \varepsilon_t, \qquad t = 1 \dots T \qquad (1)$$

Where $y_t$ is the target series, $h$ represents the number of steps ahead to forecast, $\varphi_i$ represents the $i^{th}$ coefficient of the autoregressive part of the model of order $p$ and $\varepsilon_t$ is a serially uncorrelated error term with $E(\varepsilon_t) = 0, E(\varepsilon_t^2) = \sigma_\varepsilon^2, E(\varepsilon_t^4) < \infty$, such that $E(\varepsilon_t|y_{t-i}) = 0$.

Let us also suppose that a large number (M) of indicators $X_t$, are available. In general we refer to all the Socio-Semantic Indicators (SSI) presented in Sections 3 and 3.1, considered with their lags. Given the high dimension of M, that is $M > T$, the series $X_t$ cannot be included in the model separately. However, the objective remains to extract useful information from $X_t$ in order to improve the forecasting ability of (1). This task can be accomplished by reducing the number of regressors. A standard solution to this problem is imposing a factor structure to the predictors, in order to extract a small number of components from a large set of variables, so that the relevant estimation model can be reformulated as a factor augmented autoregressive model (FAAR):

$$y_t = \sum_{i=1}^{p} \phi_i y_{t-h-i} + \xi_j' F_{t-h} + \eta_t, \quad t = 1, \dots, T \qquad (2)$$

where $F_t$ represents a $R \times 1$ vector of factors and $\xi$ a coefficient vector.

Put it differently, the *h-step*-ahead forecast is given by the following equation:

$$\hat{y}_{T+h}^{FAAR} = \sum_{i=1}^{p} \hat{\phi}_i y_{T+1-p} + \widehat{\xi'_j} \hat{F}_T \qquad (3)$$

To estimate the forecasting model (3) we followed a three-steps algorithm as suggested by Girardi, Guardabascio and Ventura [82]. The Factor Augmented Autoregressive Model (FAAR) is obtained estimating the Autoregressive Model (AR) and subsequently constructing the Factor Model (FM) on SSI indicators, to capture possible useful information not included in the AR model. Lastly, the forecasting equation is built by augmenting the AR model with the factors obtained in the second step. In particular, $F_t$ are computed by using Partial Least Squares (PLS) [83] between $y_t$ and $X_t$. Differently from Principal Components, PLS incorporates information from both the target variable and the predictors, for the definition of scores and loadings. In this regard de Jong [84] shows that the scores and loadings can be chosen in a way which describes as much as possible of covariance between the dependent variable and the regressors. We implemented the PLS algorithm on the residuals obtained at the first step mentioned above. The idea is that the residuals contain part of $y_t$ which is unexplained, thus, we tried to add information to the explanatory variables by means of the PLS applied on the SSI indicators. Moreover, the orthogonality between the residuals and the hard indicators preserves the orthogonality between the factors and the AR component.

### 3.2.1. Forecasting Procedure

All the variables included in our models were recorded on a monthly basis. The time span covered the period from 2007:1 – 2016:12. From a preliminary analysis all the variables, except for Betweenness Group Centrality and Average Response Time, showed the presence of unit roots; therefore, they were transformed with a first difference filter in order to achieve stationarity. Moreover, when applying PLS, predictors were standardized, subtracting the mean and dividing by the standard deviation. Model estimation was carried out using a rolling window out-of-sample forecasting approach which fixes a constant sample size for the in-sample regression so that, at each step, distant observations are discarded and recent ones are added. To put it in other words, an initial sample of data from t = 1, … , T is used to estimate the models and to form h-step ahead out-of-sample forecasts.

Subsequently the window is recursively moved ahead of one time period and the models re-estimated – using data from t = 2, … , T + 1. T is the window size. Both the selection of the number of lags *p* in model (1) and the optimal number of factors were defined dynamically (at each step of the rolling window of size 60 months): the first considering the Bayesian Information Criterion (BIC); the second looking at those linear combinations which explained at least 20% of the covariance between the residuals of model (1) and the variables in $X_t$. The forecast exercise is in pseudo-real-time, with an evaluation sample going from June 2013 to December 2016. The maximum value of *p* is set equal to 13, while the maximum number of factors equal to 10. This modelling approach is called adaptive and is opposed to other non-adaptive models, where the estimation of the parameters is updated without changing the equation specification, or where parameters are estimated just once and used for predictions over the entire forecasting horizon.

We evaluated model (3) (FAAR) using a first-order autoregressive model as a naïve benchmark specification, where the optimal lag of length *p* is chosen adaptively through the BIC. This means that the univariate benchmark is provided by model (1). Following, the most recent literature which proved the predictive ability of the Google Flight indicator (GF), we considered also other benchmarks: a bridge model including, together with the AR component, a certain number of lags of the GF variable, again dynamically selected through BIC (model named BRIDGE-GF):

$$\hat{y}_{T+h}^{BRIDGE-GF} = \sum_{i=1}^{p} \hat{\phi}_i y_{T+1-p} + \sum_{k=0}^{q} \hat{\gamma}_k GF_{T-k} \qquad (4)$$

and a Factor Augmented Bridge Model (FABM-GF) in which the factor is constructed from the error provided by the previous bridge model (BRIDGE-GF). This last model comprises the autoregressive terms, together with the information provided by Google Flight and the SSI indicators.

$$\hat{y}_{T+h}^{FABM-GF} = \sum_{i=1}^{p} \hat{\phi}_i y_{T+1-p} + \sum_{k=1}^{q} \hat{\gamma}_k GF_{T-k} + \hat{\xi}_j' \hat{F}_T \qquad (5)$$

To compare the forecasting performance of the different models, we referred to the mean square forecast error:

$$MSFE = \frac{\sum_{j=1}^{n}(y_{t+j}^{h} - \hat{y}_{t+j}^{h})^2}{n}$$

where $n$ is the number of months in the forecast sample and $h=1,3,6,12$. Finally, we found the set of models that forecasted equally well, relying on the model confidence set analysis of Hansen, Lunde and Nason [85]. The test for the null hypothesis of equal predictive ability at the 10% significance level was implemented using a block bootstrap scheme with 5000 resamples.

## 4. Results

This research was conceived with the idea of exploring the social dynamics of communities in online travel forums, to find new variables that could help in the prediction of international tourist arrivals in major European cities airports. Figure 2 shows the time series of the international airport arrivals for each city. We can notice a similar seasonality and an often positive time trend.

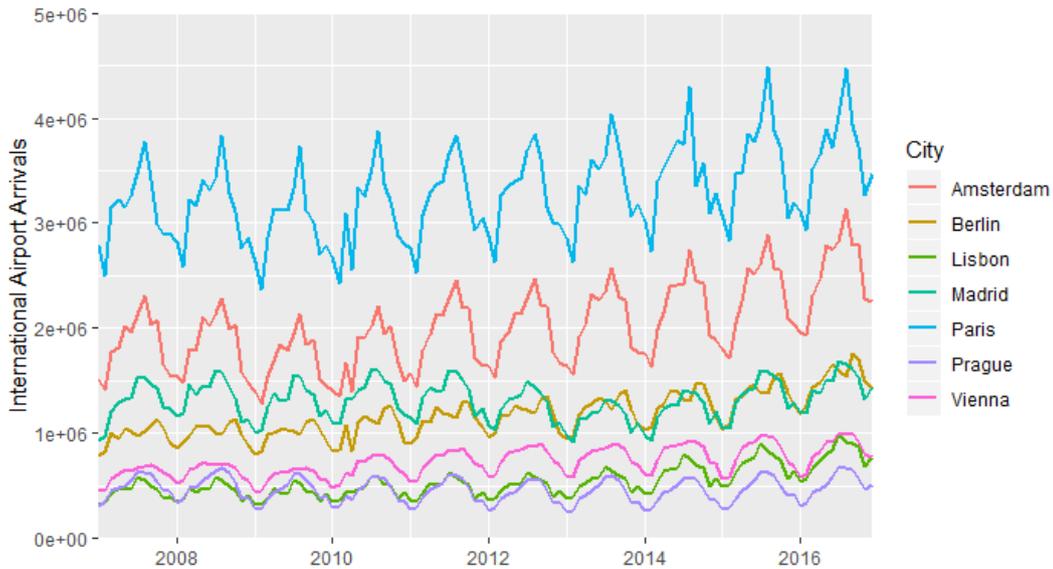

**Figure 2**. Time Series of International Airport Arrivals.

With regard to the gender distribution of users, men were predominant in almost all forums, except for Paris. As regards users' age, we see that the majority of community members were between 35 and 64, with average ages varying probably depending on the tourist attractions of each city. We think that a high average age can reflect a tendency of young users to look for tourism information using other online sources – such as Twitter, Facebook groups or Google Maps for transportations. However we cannot exclude the possibility that some young users read the forums without posting. In addition, it was not a surprise to see that people who provided tips and comments mostly lived in those very cities [10].

Even if formally modeling the major topics discussed in each forum was not in the objective of our research, we could notice that in general users were asking information about restaurants, hotels and tourist attractions (such as museums). One of the most recurrent topics was about local means of transport, with an associated negative sentiment. Sentiment of opinions about restaurants was generally more positive than comments about hotels. The museums discourse had generally positive feelings, except when discussing ticket prices and the time spent in entrance queues.

Table 3 shows the overall mean and standard deviation scores of our variables. In addition, the table reports a preliminary analysis with the results of the Dumitrescu & Hurlin test [86]. We performed this test after the removal of seasonal components, to check if our variables could granger-cause tourist arrivals up to three months in advance. Seasonal components were removed using the "STL" package in R, i.e. with a procedure based on the loess smoother [87].

| Variable | M | SD | Lag 1 Z-bar | Lag 2 Z-bar | Lag 3 Z-bar |
|---|---|---|---|---|---|
| **International Arrivals** | 1,356,743 | 967,407 | - | - | - |
| **Users Photos** | 121,718.30 | 220,515.00 | 5.24*** | 14.79*** | 11.93*** |
| **Users Level** | 9,282.06 | 1,3350.22 | 56.81*** | 54.77*** | 37.95*** |
| **Percentage Male** | .64 | .12 | 3.77*** | 2.80** | 3.22** |
| **Average Age** | 48.37 | 3.90 | 2.70* | 2.98** | 4.06*** |
| **Activity** | 3,227.13 | 4,671.93 | 58.00*** | 55.73*** | 40.64*** |
| **Group Betweenness Centrality** | .33 | .09 | -0.40 | 0.95 | 3.08** |
| **Sentiment** | .50 | .03 | 3.24** | 2.05* | 2.18** |
| **Complexity** | 5.64 | .40 | 51.38*** | 32.57*** | 39.91*** |
| **Group Degree Centrality** | .44 | .11 | 8.25*** | 10.69*** | 8.04*** |
| **Average Response Time** | 6.97 | 3.55 | 1.09 | 0.60 | 2.64** |
| **Rotating Leadership** | 18.04 | 3.11 | 9.00*** | 9.07*** | 3.64*** |
| **New Users** | 76.80 | 104.54 | 30.31*** | 33.16*** | 21.81*** |
| **Google Trend Flights** | 27.65 | 21.04 | 13.59*** | 19.52*** | 21.43*** |
| **Google Trend Holidays** | 33.13 | 21.04 | 5.26*** | 14.40*** | 10.03*** |

*p < .05; **p < .01; ***p < .001.

**Table 3**. Descriptive Statistics and Dumitrescu & Hurlin Test

Tests results show a potentially significant association of international airport arrivals with all our social network and semantic metrics, apart from average response time and group betweenness centrality at lags 1 and 2. User level, activity and complexity are the variables that exhibit the strongest associations with the international arrivals.

Table 4 shows the results of the different forecasting models we presented in Section 3, i.e. the benchmark represented by the Autoregre*s*sive Model (AR), the Factor Augment Autoregressive Model (FAAR) complementing AR with information from the SSI indicators, the Bridge Model which includes information from Google Flight together with the AR component (BRIDGE –GF), and the Factor Augmented Bridge Model (FABM-GF) which is BRIDGE-GF augmented with the Factor coming from SSI indicators. The average Mean Square Forecast Error (MSFE) of each model is reported in the table as a ratio to the average MSFE of the univariate autoregressive model (AR), for four different forecasting horizons. Accordingly, the performance of one model is better than the AR if the corresponding ratio is lower than 1. In the table, we also report the average Root Mean Squared Error, for each model and forecasting horizon.
Asterisks in the table are used to mark those models which are included in the superior set at the 10% significance level.

Table 4 shows that information coming from semantic and social network variables can significantly improve the forecasting performance of international airport arrivals provided by AR or BRIDGE-GF models. Indeed, models which included these new variables were the best choice in 79% of cases and 93% of times were included in the superior set. The models with the AR component and our predictors (FAAR), without Google Trend Flights, could outperform the other models in 54% of cases. Worst performance was obtained for six-month forecasts. At this horizon, AR models represented a better choice for 5 cities out of 7 (Lisbon, Madrid, Paris, Prague and Vienna), even if FAAR models were still included in the superior set, for all these cities apart from Prague. On the other hand, our predictors led to better forecasts, even at h = 6, for Amsterdam and Berlin. We only used the Google Trend Flights indicator as this always had a better performance than Google Trend Holidays. In general the inclusion of this last variable in the models did not lead to better results. We found a potential collinearity issue – due to the high correlation of activity, user level and new contacts – which was however efficiently handled by our factor models.

In order to better evaluate the robustness of our models, we tested several other approaches and combinations of predictors, to see whether good forecasts could be obtained using simpler metrics, without the need of calculating SSI indicators. In particular, we tried to understand if the metric of activity could be substituted by two simpler metrics: the number of posts and the average number of replies per thread. Similarly, we considered the heterogeneity of user levels – i.e. their standard deviation – as a possible replacement for group degree and group betweenness centrality, which measure heterogeneity in user centralities. We wanted to be sure that the effort put in the calculation of SSI indicators had a reason. Consistently, Table 4 also shows the results of a model named "BRIGDE-OTH-GF", which includes the AR component, the Google Trend Flights indicator and the three simpler variables we just mentioned. Results show that this simpler model was never the best choice – except for Madrid at h=3, where the model was included, as all others, in the confidence set. In general, the use of number of posts, average number of replies per thread and heterogeneity of user levels led to significantly worse forecasts. In 26 cases out of 28, these models were outperformed by models which included our SSI indicators (FAAR and FABM-GF). Similarly, the performance of BRIDGE-GF, was almost never improved by the inclusion of the three variables. These alternative metrics gave little contribution to the information already captured by the AR component and by the Google Flight indicator. Lastly, we tried replacing activity and centrality metrics in our FAAR and FABM-GF models. This led to no better results.

| Forecasting Horizon | Model | Amsterdam Rel. MSE | Amsterdam RMSE | Berlin Rel. MSE | Berlin RMSE | Lisbon Rel. MSE | Lisbon RMSE | Madrid Rel. MSE | Madrid RMSE | Paris Rel. MSE | Paris RMSE | Prague Rel. MSE | Prague RMSE | Vienna Rel. MSE | Vienna RMSE |
|---|---|---|---|---|---|---|---|---|---|---|---|---|---|---|---|
| One Month | AR | 1.00 | 75747 | 1.00 | 51568 | 1.00* | 27040 | 1.00 | 48758 | 1.00 | 20788 | 1.00* | 19622 | 1.00 | 30439 |
|  | FAAR | 0.6791 | 62421 | **0.8015*** | **46167** | 0.9017* | 25677 | 0.9174 | 46702 | **0.5196*** | 14985 | 0.9555* | **19180** | 0.7929 | 27105 |
|  | FABM-GF | **0.6627*** | 61662 | 0.8518* | 47592 | **0.8882*** | **25483** | 0.7878* | 43276 | 0.529* | **15120** | 0.9832* | 19456 | **0.7668*** | **26655** |
|  | BRIGDE-OTH-GF | 1.2757 | 85555 | 1.3798 | 60575 | 0.9765* | 26721 | 1.0506* | 49977 | 0.8999 | 19719 | 1.2458 | 21901 | 1.0347 | 30962 |
|  | BRIDGE-GF | 1.0231 | 76616 | 1.0471 | 52769 | 1.0256* | 27383 | 0.8514* | 44989 | 0.9818 | 20598 | 1.0292* | 19906 | 0.9586 | 29802 |
| Three Months | AR | 1.00 | 74638 | 1.00* | 52710 | 1.00 | 25804 | 1.00* | 46475 | 1.00* | 19875 | 1.00* | 19818 | 1.00* | 29245 |
|  | FAAR | **0.9855*** | **74096** | **0.9722*** | **51972** | 1.0192 | 26050 | 0.9873* | 46180 | **0.9694*** | **19568** | **0.9821*** | **19640** | **0.9948*** | **29169** |
|  | FABM-GF | 1.1384 | 79636 | 1.156 | 56672 | 0.9593* | 25273 | 1.0278* | 47117 | 0.9704* | 19578 | 1.0117 | 19934 | 1.0374 | 29787 |
|  | BRIGDE-OTH-GF | 2.0090 | 10579 | 1.3333 | 60685 | 1.0645* | 26622 | **0.9465*** | **45216** | 1.0410* | 20278 | 1.1848 | 21572 | 1.1598 | 31496 |
|  | BRIDGE-GF | 1.1446 | 79852 | 1.1621 | 56822 | **0.9535*** | **25197** | 1.0411* | 47421 | 0.983* | 19705 | 1.0341 | 20153 | 1.0448 | 29893 |
| Six Months | AR | 1.00 | 73070 | 1.00* | 51126 | **1.00*** | **26366** | **1.00*** | **48637** | **1.00*** | **19091** | **1.00*** | **20046** | **1.00*** | **29515** |
|  | FAAR | 0.927* | 70354 | **0.9557*** | **49980** | 1.0016* | 26386 | 1.029* | 49336 | 1.0258* | 19335 | 1.0422 | 20464 | 1.0076* | 29627 |
|  | FABM-GF | **0.8427*** | **67076** | 1.1328 | 54414 | 1.1889 | 28749 | 1.1058 | 51144 | 1.063* | 19683 | 1.0575 | 20614 | 1.1607* | 31798 |
|  | BRIGDE-OTH-GF | 1.6248 | 93141 | 1.2757 | 57746 | 1.5074 | 32370 | 1.2617 | 54632 | 1.1637 | 20595 | 1.0120 | 20166 | 1.3249 | 33974 |
|  | BRIDGE-GF | 0.9459 | 71066 | 1.1808 | 5557 | 1.1783 | 28619 | 1.0819 | 50589 | 1.0661* | 19712 | 1.0113 | 20159 | 1.1514* | 31672 |
| One Year | AR | 1.00 | 73128 | 1.00* | 48888 | 1.00 | 28990 | 1.00 | 52373 | 1.00* | 22837 | 1.00 | 23778 | 1.00 | 31304 |
|  | FAAR | **0.7258*** | **62299** | 0.9382* | 47352 | 0.6481* | 23339 | **0.8226*** | **47500** | 0.9546* | **22313** | 0.5999* | **18417** | **0.8651*** | **29116** |
|  | FABM-GF | 0.8292 | 66592 | **0.9339*** | **47244** | **0.6471*** | **23321** | 0.8772* | 49051 | 1.0593 | 23504 | 0.6002* | 18422 | 0.945 | 30430 |
|  | BRIGDE-OTH-GF | 2.2544 | 109800 | 1.4894 | 59662 | 0.9557 | 28340 | 1.0905 | 54692 | 1.1808 | 24816 | 1.1970 | 26015 | 1.1490 | 33556 |
|  | BRIDGE-GF | 1.1464 | 78300 | 0.9780* | 48348 | 1.0192 | 29268 | 1.0493 | 53649 | 1.0987 | 23938 | 1.0736 | 24637 | 1.0776 | 32496 |

*Note.* The best result for each forecasting horizon (h) is shown in bold. * indicates models that belong to the superior set at the 10% level. RMSE is the average Root Mean Squared Error of forecasts. Rel. MSE is the ratio of Mean Squared Errors of each model with respect to its corresponding AR model.

**Table 4**. Accuracy of Forecasting Models

Table 5 shows the analysis of the weights determined by the FAAR model for each social network and semantic variable. Most important predictors are those which more frequently obtained a coefficient high enough to be in the IV quartile of the weight distribution of the factor selected at each step of each rolling window. More in detail, as the evaluation sample for h=1 counts 46 observations, Table 4 summarizes information coming from 322 weight distributions (7 cities multiplied by 46 evaluations). For each distribution, we count how many times the weight of the different indicators belongs to a specific quartile. Group betweenness and degree centrality and language complexity are those variables which appear more often in the IV quartile, meaning that more frequently they have a bigger role in influencing the forecasts. Results – which are provided in the table for a forecasting horizon of one month – were consistent and stable across all the other horizons. The only exception was for new contacts, which becomes more important for one-year forecasts (54.76% of times in the IV quartile), and rotating leadership, which had an opposite trend (38.10% of times in the IV quartile for one-year forecasts).

| Variable | II quartile | III quartile | IV quartile |
|---|---|---|---|
| **Users Photo** | 14.29% | 21.43% | 45.24% |
| **Users Level** | 9.52% | 19.05% | 33.33% |
| **Percentage Male** | 0.00% | 14.29% | **57.14%** |
| **Average Age** | 7.14% | 9.52% | 50.00% |
| **Activity** | 7.14% | 11.90% | 38.10% |
| **Group Betweenness Centrality** | 4.76% | 9.52% | **69.05%** |
| **Sentiment** | 11.90% | 16.67% | 42.86% |
| **Complexity** | 7.69% | 7.69% | **61.54%** |
| **Group Degree Centrality** | 4.76% | 19.05% | **57.14%** |
| **Average Response Time** | 9.52% | 21.43% | 35.71% |
| **Rotating Leadesrship** | 11.90% | 19.05% | 50.00% |
| **New Users** | 16.67% | 14.29% | 47.62% |

**Table 5**. Analysis of FAAR weights.

## 5. Discussion, Limitations and Future Research

The level of tourism impacts the economy of a country [88]. Forecasting tourism demand has important implications for policy makers, company managers working in the tourism industry and several other stakeholders. At the same time, information and opinions exchanged among tourists can influence the number of visitors to specific destinations or attractions and the image formation for places tourists have not yet visited [89,90]. In our case study, we present a ten-year analysis of the TripAdvisor travel forum, carried out by developing a specific web crawler and combining methods and tools from social network and semantic analysis. Descriptive statistics indicate that male users were predominant, and that majority of forum participants were between 35 and 64 years old. Among the most recurring topics, we found requests for information about local means of transport, associated with an average lower sentiment. This is also due to the fact that users' comments about local means of transport are commonly shared in forums and not in separate reviews – as it can happen for hotels, airlines, museums and restaurants.

Our findings indicate that variables coming from the analysis of online forums can significantly improve the forecasting models which consider the volume of online search queries (measured by the Google Trend index). Social network and sematic variables could improve forecasts in 79% of cases and models including them were in the superior set in 93% of cases. There were exceptions to this improved performance at six-month forecasting horizon, where for 5 cities our predictors could not improve the results of AR models.

Overall, forum language complexity and the centralization of communication (group degree and betweenness centrality) emerged as the most important predictors. Higher complexity seems to anticipate more arrivals. It can be indicative of a more informative language [19], as this measure is higher when new words are used in the forum posts. Therefore, one explanation of the link between complexity and arrivals could be that prospective tourists look for information about their destinations, posting questions which demand for new knowledge; answers to these questions bring new content in the forums, making the language more complex. Similarly, higher centralization of online interactions, can be a signal of the presence of eminent contributors, i.e. informal moderators who are probably local experts who share their knowledge with prospective tourists.

The percentage of male users could also contribute to the improvement of forecasting performance. The number of forum posts (activity), on the other hand, was less informative than the number of new users joining the forum. Even if significant in our preliminary analysis of granger causality, other variables – such as average response time and users level – were less important to forecasting purposes. Rotating leadership, which seems to have a pivotal role for online community growth [60], had high model weights in 50% of cases, with a declining trend for longer forecasting horizons. It seems that the presence of local experts who dominate conversations can support prospective tourists more than democracy in interactions and plurality of opinions.

In general, we maintain that traditional models for the prediction of tourism demand – based on more conventional metrics, such as the univariate analysis of tourist arrivals – can be improved by extracting big data from online sources. In this sense, the variables that we presented in this study have a potential which should be studied more, using different forecasting techniques. In addition, the theoretical reasons behind the different contribution of each of them could be investigated further, in dedicated future research.

This study has also other limitations which we plan to address in future studies. The choice of studying the number of international airport arrivals as the dependent variable does not consider visitors entering a country by other means of transport; this variable is also not suitable to identify people traveling for work. Moreover, a part of international travelers might land to a city airport and then move to other places. As already discussed in Section 3, the selection of other measures would imply other limitations. For example, if we count the number of visa requirements, we would need to take into account that citizens of the European Union do not need a visa to travel in EU countries. Similarly, counting the number of nights spent in accommodation establishments is not always a good proxy for the number of tourists [70]. In general, we remind the reader that the main objective of this study is to give evidence to the potential of new variables which can be useful for tourist arrival predictions, and which can be relatively easy to extract from the web. Accordingly, even if international airport arrivals is not the perfect proxy of tourism demand, we maintain this is a reasonable indicator for the purpose of our research. This choice is also consistent with previous research [70,73]. We advocate future research to study online travel communities interacting on Facebook groups or on other social media platforms – where the number of users younger than 35 years old is potentially larger. Could SSI variables measured on other online platforms be more informative than those measured on the TripAdvisor travel forum? It would also be interesting to study the effects of language sentiment and complexity considering languages other than English. Scholars might want to test additional dependent variables, or forecast tourism demand at different levels (for example analyzing online communities to predict the number of visitors to specific museums or attractions). We advocate future research to assess the predictive power of SSI indicators considering cities outside Europe and smaller, less popular, cities which are not capitals. The development of a more resilient and flexible crawler is also in our future plans, in order to reduce/eliminate data quality issues and examine interactions taking place on other web platforms. Lastly, in-depth analysis of the major topics of each city forum could provide further insights.

## 6. Conclusions

The use of big data in tourism creates new challenges [91]. We show that applying social network and semantic analysis to big data extracted from online travel forums can help making predictions of international tourist arrivals. Our metrics prove their value in increasing the forecasting accuracy of models which consider the volume of online search queries. This research findings contribute to the research about tourism forecasting, presenting a new approach and new metrics. Past research mostly considered other sources of online data – such as web search queries or online reviews [22,24,77–79], whereas interaction dynamics in travel forums were less explored. Moreover, the use of social network analysis in tourism is recent and new [54].

This research has practical implications for researchers, policy makers and business managers working in the tourism industry – who could, for example, adjust prices or make more accurate sales forecasts. Similarly to Song and Witt [92], we maintain that accurate forecasts are vital for: efficient planning of tourism-related businesses, dealing with extremely perishable products (to avoid for example overbookings or empty hotel rooms); adequate appraisal of public projects and planning of investments in destination infrastructures; appropriate support of governmental decision-making processes, which regard the allocation of resources and the formulation of medium-to-long term tourism strategies.

**Acknowledgements**

The authors are grateful to the Advisory Committee on Statistical Methods of the Italian National Institute of Statistics and to Claudia Colladon, for their help in revising this manuscript. The authors are also grateful to the reviewers for their thoughtful and precious advice.

**References**

[1]  B. McKercher, Towards a taxonomy of tourism products, Tourism Management. 54 (2016) 196–208. doi:10.1016/j.tourman.2015.11.008.

[2]  G. Zhu, K.K.F. So, S. Hudson, Inside the sharing economy: Understanding consumer motivations behind the adoption of mobile applications, International Journal of Contemporary Hospitality Management. 29 (2017) 2218–2239. doi:10.1108/IJCHM-09-2016-0496.

[3]  Z. Xiang, V.P. Magnini, D.R. Fesenmaier, Information technology and consumer behavior in travel and tourism: Insights from travel planning using the internet, Journal of Retailing and Consumer Services. 22 (2015) 244–249. doi:10.1016/j.jretconser.2014.08.005.

[4]  R. Law, R. Leung, D. Buhalis, Information technology applications in hospitality and tourism: A review of publications from 2005 to 2007, Journal of Travel and Tourism Marketing. 26 (2009) 599–623. doi:10.1080/10548400903163160.

[5]  Y. Li, C. Hu, C. Huang, L. Duan, The concept of smart tourism in the context of tourism information services, Tourism Management. 58 (2017) 293–300. doi:10.1016/j.tourman.2016.03.014.

[6]  G. Prayag, S. Hosany, K. Odeh, The role of tourists' emotional experiences and satisfaction in understanding behavioral intentions, Journal of Destination Marketing and Management. 2 (2013) 118–127. doi:10.1016/j.jdmm.2013.05.001.

[7]  Z. Xiang, Z. Schwartz, J.H. Gerdes, M. Uysal, What can big data and text analytics tell us about hotel guest experience and satisfaction?, International Journal of Hospitality Management. 44 (2015) 120–130. doi:10.1016/j.ijhm.2014.10.013.

[8]  T. Hennig-Thurau, K.P. Gwinner, G. Walsh, D.D. Gremler, Electronic word-of-mouth via consumer-opinion platforms: What motivates consumers to articulate themselves on the Internet?, Journal of Interactive Marketing. 18 (2004) 38–52. doi:10.1002/dir.10073.

[9]  J.Y.C. Ho, M. Dempsey, Viral marketing: Motivations to forward online content, Journal of

Business Research. 63 (2010) 1000–1006. doi:10.1016/j.jbusres.2008.08.010.

[10] J. Miguéns, R. Baggio, C. Costa, Social media and Tourism Destinations: TripAdvisor Case Study, Advances in Tourism Research. 26 (2008) 26–28. doi:10.1088/1751-8113/44/8/085201.

[11] S.P. Eslami, M. Ghasemaghaei, K. Hassanein, Which online reviews do consumers find most helpful? A multi-method investigation, Decision Support Systems. 113 (2018) 32–42. doi:10.1016/j.dss.2018.06.012.

[12] M. Farhadloo, R.A. Patterson, E. Rolland, Modeling customer satisfaction from unstructured data using a Bayesian approach, Decision Support Systems. 90 (2016) 1–11. doi:10.1016/j.dss.2016.06.010.

[13] G. Li, R. Law, H.Q. Vu, J. Rong, X. (Roy) Zhao, Identifying emerging hotel preferences using Emerging Pattern Mining technique, Tourism Management. 46 (2015) 311–321. doi:10.1016/j.tourman.2014.06.015.

[14] G. George, M.R. Haas, A. Pentland, Big Data and Management, Academy of Management Journal. 57 (2014) 321–326. doi:10.5465/amj.2014.4002.

[15] A. Gandomi, M. Haider, Beyond the hype: Big data concepts, methods, and analytics, International Journal of Information Management. 35 (2015) 137–144. doi:10.1016/j.ijinfomgt.2014.10.007.

[16] R.M. Chang, R.J. Kauffman, Y. Kwon, Understanding the paradigm shift to computational social science in the presence of big data, Decision Support Systems. 63 (2014) 67–80. doi:10.1016/j.dss.2013.08.008.

[17] F.E.A. Horita, J.P. de Albuquerque, V. Marchezini, E.M. Mendiondo, Bridging the gap between decision-making and emerging big data sources: An application of a model-based framework to disaster management in Brazil, Decision Support Systems. 97 (2017) 12–22. doi:10.1016/j.dss.2017.03.001.

[18] S. Wasserman, K. Faust, Social Network Analysis: Methods and Applications, Cambridge

University Press, New York, NY, 1994. doi:10.1525/ae.1997.24.1.219.

[19]   P.A. Gloor, Sociometrics and Human Relationships: Analyzing Social Networks to Manage Brands, Predict Trends, and Improve Organizational Performance, Emerald Publishing Limited, London, UK, 2017.

[20]   H. Song, G. Li, Tourism demand modelling and forecasting - A review of recent research, Tourism Management. 29 (2008) 203–220. doi:10.1016/j.tourman.2007.07.016.

[21]   C. Lim, M. McAleer, Forecasting tourist arrivals, Annals of Tourism Research. 28 (2001) 965–977. doi:10.1016/S0160-7383(01)00006-8.

[22]   P.F. Bangwayo-Skeete, R.W. Skeete, Can Google data improve the forecasting performance of tourist arrivals? Mixed-data sampling approach, Tourism Management. 46 (2015) 454–464. doi:10.1016/j.tourman.2014.07.014.

[23]   F. Kallasidis, Web Search activity: A Forecast Tool for Tourist Arrivals in Cyprus., 2015. https://repository.ihu.edu.gr/xmlui/handle/11544/14480.

[24]   B.A. Sparks, V. Browning, The impact of online reviews on hotel booking intentions and perception of trust, Tourism Management. 32 (2011) 1310–1323. doi:10.1016/j.tourman.2010.12.011.

[25]   J.M. Alcántara-Pilar, S. del Barrio-García, E. Crespo-Almendros, L. Porcu, Toward an understanding of online information processing in e-tourism: does national culture matter?, Journal of Travel & Tourism Marketing. (2017) 1–15. doi:10.1080/10548408.2017.1326363.

[26]   M. Sigala, K. Chalkiti, Investigating the exploitation of web 2.0 for knowledge management in the Greek tourism industry: An utilisation-importance analysis, Computers in Human Behavior. 30 (2014) 800–812. doi:10.1016/j.chb.2013.05.032.

[27]   U. Gretzel, M. Sigala, Z. Xiang, C. Koo, Smart tourism: foundations and developments, Electronic Markets. 25 (2015) 179–188. doi:10.1007/s12525-015-0196-8.

[28]   J. Lu, Z. Mao, M. Wang, L. Hu, Goodbye maps, hello apps? Exploring the influential

determinants of travel app adoption, Current Issues in Tourism. 18 (2015) 1059–1079. doi:10.1080/13683500.2015.1043248.

[29]   Y. Narangajavana, L.J. Callarisa Fiol, M.Á. Moliner Tena, R.M. Rodríguez Artola, J. Sánchez García, The influence of social media in creating expectations. An empirical study for a tourist destination, Annals of Tourism Research. 65 (2017) 60–70. doi:10.1016/j.annals.2017.05.002.

[30]   S. Amaro, P. Duarte, C. Henriques, Travelers' use of social media: A clustering approach, Annals of Tourism Research. 59 (2016) 1–15. doi:10.1016/j.annals.2016.03.007.

[31]   M. Fuchs, W. Höpken, M. Lexhagen, Big data analytics for knowledge generation in tourism destinations - A case from Sweden, Journal of Destination Marketing and Management. 3 (2014) 198–209. doi:10.1016/j.jdmm.2014.08.002.

[32]   S.J. Miah, H.Q. Vu, J. Gammack, M. McGrath, A Big Data Analytics Method for Tourist Behaviour Analysis, Information and Management. 54 (2017) 771–785. doi:10.1016/j.im.2016.11.011.

[33]   X. Yang, B. Pan, J.A. Evans, B. Lv, Forecasting Chinese tourist volume with search engine data, Tourism Management. 46 (2015) 386–397. doi:10.1016/j.tourman.2014.07.019.

[34]   D. Gavilan, M. Avello, G. Martinez-Navarro, The influence of online ratings and reviews on hotel booking consideration, Tourism Management. 66 (2018) 53–61. doi:10.1016/j.tourman.2017.10.018.

[35]   B. Zeng, R. Gerritsen, What do we know about social media in tourism? A review, Tourism Management Perspectives. 10 (2014) 27–36. doi:10.1016/j.tmp.2014.01.001.

[36]   Y.H. Hwang, D. Jani, H.K. Jeong, Analyzing international tourists' functional information needs: A comparative analysis of inquiries in an on-line travel forum, Journal of Business Research. 66 (2013) 700–705. doi:10.1016/j.jbusres.2011.09.006.

[37]   A. Hirashima, J. Jones, C.S. Bonham, P. Fuleky, Forecasting in a Mixed Up World: Nowcasting Hawaii Tourism, Annals of Tourism Research. 63 (2017) 191–202.

doi:10.1016/j.annals.2017.01.007.

[38]   H. Hassani, E.S. Silva, N. Antonakakis, G. Filis, R. Gupta, Forecasting accuracy evaluation of tourist arrivals, Annals of Tourism Research. 63 (2017) 112–127. doi:10.1016/j.annals.2017.01.008.

[39]   Z. Cao, G. Li, H. Song, Modelling the interdependence of tourism demand: The global vector autoregressive approach, Annals of Tourism Research. 67 (2017) 1–13. doi:10.1016/j.annals.2017.07.019.

[40]   J.K.S. Jacobsen, A.M. Munar, Tourist information search and destination choice in a digital age, Tourism Management Perspectives. 1 (2012) 39–47. doi:10.1016/j.tmp.2011.12.005.

[41]   G. Athanasopoulos, A. de Silva, Multivariate Exponential Smoothing for Forecasting Tourist Arrivals, Journal of Travel Research. 51 (2012) 640–652. doi:10.1177/0047287511434115.

[42]   U. Gunter, I. Önder, Forecasting international city tourism demand for Paris: Accuracy of uni- and multivariate models employing monthly data, Tourism Management. 46 (2015) 123–135. doi:10.1016/j.tourman.2014.06.017.

[43]   Y. Yang, B. Pan, H. Song, Predicting Hotel Demand Using Destination Marketing Organization's Web Traffic Data, Journal of Travel Research. 53 (2014) 433–447. doi:10.1177/0047287513500391.

[44]   B. Pan, D. Chenguang Wu, H. Song, Forecasting hotel room demand using search engine data, Journal of Hospitality and Tourism Technology. 3 (2012) 196–210. doi:10.1108/17579881211264486.

[45]   X. Li, B. Pan, R. Law, X. Huang, Forecasting tourism demand with composite search index, Tourism Management. 59 (2017) 57–66. doi:10.1016/j.tourman.2016.07.005.

[46]   C. Artola, F. Pinto, P. de Pedraza García, Can internet searches forecast tourism inflows?, International Journal of Manpower. 36 (2015) 103. doi:10.1108/IJM-12-2014-0259.

[47]   H. Choi, H. Varian, Predicting the Present with Google Trends, Economic Record. 88 (2012) 2–

9. doi:10.1111/j.1475-4932.2012.00809.x.

[48]  U. Gunter, I. Önder, Forecasting city arrivals with Google Analytics, Annals of Tourism Research. 61 (2016) 199–212. doi:10.1016/j.annals.2016.10.007.

[49]  S. Shen, G. Li, H. Song, An Assessment of Combining Tourism Demand Forecasts over Different Time Horizons, Journal of Travel Research. 47 (2008) 197–207. doi:10.1177/0047287508321199.

[50]  X. Sun, W. Sun, J. Wang, Y. Zhang, Y. Gao, Using a Grey-Markov model optimized by Cuckoo search algorithm to forecast the annual foreign tourist arrivals to China, Tourism Management. 52 (2016) 369–379. doi:10.1016/j.tourman.2015.07.005.

[51]  E.D. Höpken W. Fuchs M., Kronenberg K., Lexhagen M., Big Data as Input for Predicting Tourist Arrivals, Annals of Tourism Research. 28 (2017) 1070–1072. doi:10.1016/S0160-7383(01)00012-3.

[52]  S. Shen, G. Li, H. Song, Combination forecasts of International tourism demand, Annals of Tourism Research. 38 (2011) 72–89. doi:10.1016/j.annals.2010.05.003.

[53]  H. Dali, M. Yutaka, Predicting tourism trends through the data of online communications, in: 29th Annual Conference of the Japanese Society for Artificial Intelligence, 2015: pp. 1–4.

[54]  C. Casanueva, Á. Gallego, M.-R. García-Sánchez, Social network analysis in tourism, Current Issues in Tourism. 3500 (2014) 1–20. doi:10.1080/13683500.2014.990422.

[55]  V. Browning, K.K.F. So, B. Sparks, The Influence of Online Reviews on Consumers' Attributions of Service Quality and Control for Service Standards in Hotels, Journal of Travel & Tourism Marketing. 30 (2013) 23–40. doi:10.1080/10548408.2013.750971.

[56]  Q. Ye, R. Law, B. Gu, W. Chen, The influence of user-generated content on traveler behavior: An empirical investigation on the effects of e-word-of-mouth to hotel online bookings, Computers in Human Behavior. 27 (2011) 634–639. doi:10.1016/j.chb.2010.04.014.

[57]  E. Pantano, C.V. Priporas, N. Stylos, 'You will like it!' using open data to predict tourists'

response to a tourist attraction, Tourism Management. 60 (2017) 430–438. doi:10.1016/j.tourman.2016.12.020.

[58] M. Siering, A. V. Deokar, C. Janze, Disentangling consumer recommendations: Explaining and predicting airline recommendations based on online reviews, Decision Support Systems. 107 (2018) 52–63. doi:10.1016/j.dss.2018.01.002.

[59] K.-H. Yoo, U. Gretzel, Comparison of Deceptive and Truthful Travel Reviews, in: W. Höpken, U. Gretzel, R. Law (Eds.), Information and Communication Technologies in Tourism 2009, Vienna, AU, 2009: pp. 37–47. doi:10.1016/S0160-7383(01)00012-3.

[60] G. Antonacci, A. Fronzetti Colladon, A. Stefanini, P. Gloor, It is rotating leaders who build the swarm: social network determinants of growth for healthcare virtual communities of practice, Journal of Knowledge Management. 21 (2017) 1218–1239. doi:10.1108/JKM-11-2016-0504.

[61] J. Koh, Y.G. Kim, Knowledge sharing in virtual communities: an e-business perspective., Expert Systems with Applications. 26 (2004) 155–166.

[62] A. Van Looy, Social media management: technologies and strategies for creating business value, Springer, New York, NY, 2016. doi:10.1007/978-3-319-21990-5.

[63] A. Fronzetti Colladon, F. Vagaggini, Robustness and stability of enterprise intranet social networks: The impact of moderators, Information Processing & Management. 53 (2017) 1287–1298. doi:10.1016/j.ipm.2017.07.001.

[64] Y. Huang, C. Basu, M.K. Hsu, Exploring Motivations of Travel Knowledge Sharing on Social Network Sites: An Empirical Investigation of U.S. College Students, Journal of Hospitality Marketing & Management. 19 (2010) 717–734. doi:10.1080/19368623.2010.508002.

[65] M. Kang, B. Kim, P. Gloor, G.W. Bock, Understanding the effect of social networks on user behaviors in community-driven knowledge services, Journal of the American Society for Information Science and Technology. 62 (2011) 1066–1074. doi:10.1002/asi.21533.

[66] P. Gloor, A. Fronzetti Colladon, G. Giacomelli, T. Saran, F. Grippa, The impact of virtual

mirroring on customer satisfaction, Journal of Business Research. 75 (2017) 67–76. doi:10.1016/j.jbusres.2017.02.010.

[67] D. Yin, S.D. Bond, H. Zhang, Anxious or Angry? Effects of Discrete Emotions on the Perceived Helpfulness of Online Reviews1., MIS Quarterly. 38 (2014) 539–560. http://search.ebscohost.com/login.aspx?direct=true&db=a9h&AN=95756002&lang=zh-cn&site=ehost-live.

[68] M. Salehan, D.J. Kim, Predicting the performance of online consumer reviews: A sentiment mining approach to big data analytics, Decision Support Systems. 81 (2016) 30–40. doi:10.1016/j.dss.2015.10.006.

[69] R. V Kozinets, The Field Behind the Screen: Using Netnography for Marketing Research in Online Communities, Journal of Marketing Research. 39 (2002) 61–72. doi:10.1509/jmkr.39.1.61.18935.

[70] T. Garín-Muñoz, Inbound international tourism to Canary Islands: A dynamic panel data model, Tourism Management. 27 (2006) 281–291. doi:10.1016/j.tourman.2004.10.002.

[71] T. Bieger, A. Wittmer, Air transport and tourism - Perspectives and challenges for destinations, airlines and governments, Journal of Air Transport Management. 12 (2006) 40–46. doi:10.1016/j.jairtraman.2005.09.007.

[72] A. Dupeyras, N. Maccallum, Indicators for Measuring Competitiveness in Tourism: A Guidance Document, OECD Tourism Papers. 2013/02 (2013) 1–62. doi:10.1787/5k47t9q2t923-en.

[73] J.L. Eugenio-Martin, Estimating the tourism demand impact of public infrastructure investment: The case of Malaga airport expansion, Tourism Economics. 22 (2016) 254–268. doi:10.5367/te.2016.0547.

[74] S.P. Borgatti, M.G. Everett, J.C. Johnson, Analyzing Social Networks, SAGE Publications, New York, NY, 2013.

[75] Y.H. Kidane, P.A. Gloor, Correlating temporal communication patterns of the Eclipse open

source community with performance and creativity, Computational and Mathematical Organization Theory. 13 (2007) 17–27.

[76] L. Brönnimann, Analyse der Verbreitung von Innovationen in sozialen Netzwerken, 2014. http://www.twitterpolitiker.ch/documents/Master_Thesis_Lucas_Broennimann.pdf.

[77] R. Rivera, A dynamic linear model to forecast hotel registrations in Puerto Rico using Google Trends data, Tourism Management. 57 (2016) 12–20. doi:10.1016/j.tourman.2016.04.008.

[78] I. Önder, Forecasting tourism demand with Google trends: Accuracy comparison of countries versus cities, International Journal of Tourism Research. 19 (2017) 648–660. doi:10.1002/jtr.2137.

[79] M. Jackman, S. Naitram, Research Note: Nowcasting Tourist Arrivals in Barbados – Just Google it!, Tourism Economics. 21 (2015) 1309–1313. doi:10.5367/te.2014.0402.

[80] L.C. Freeman, Centrality in social networks conceptual clarification, Social Networks. 1 (1979) 215–239.

[81] L. Brönnimann, Multilanguage sentiment-analysis of Twitter data on the example of Swiss politicians, Windisch, Switzerland, 2013. http://www.twitterpolitiker.ch/Paper_Swiss_Politicians_On_Twitter.pdf.

[82] A. Girardi, B. Guardabascio, M. Ventura, Factor-Augmented Bridge Models (FABM) and Soft Indicators to Forecast Italian Industrial Production, Journal of Forecasting. 35 (2016) 542–552. doi:10.1002/for.2393.

[83] G. Cubadda, B. Guardabascio, A medium-N approach to macroeconomic forecasting, Economic Modelling. 29 (2012) 1099–1105. doi:10.1016/j.econmod.2012.03.027.

[84] S. de Jong, SIMPLS: An alternative approach to partial least squares regression, Chemometrics and Intelligent Laboratory Systems. 18 (1993) 251–263. doi:10.1016/0169-7439(93)85002-X.

[85] P.R. Hansen, A. Lunde, J.M. Nason, The Model Confidence Set, Econometrica. 79 (2011) 453–497. doi:10.3982/ECTA5771.

[86]  E.I. Dumitrescu, C. Hurlin, Testing for Granger non-causality in heterogeneous panels, Economic Modelling. 29 (2012) 1450–1460. doi:10.1016/j.econmod.2012.02.014.

[87]  R.B. Cleveland, W.S. Cleveland, J.E. McRae, I. Terpenning, STL: A seasonal-trend decomposition procedure based on loess, Journal of Official Statistics. 6 (1990) 3–73. doi:citeulike-article-id:1435502.

[88]  L. Xue, D. Kerstetter, C. Hunt, Tourism development and changing rural identity in China, Annals of Tourism Research. 66 (2017) 170–182. doi:10.1016/j.annals.2017.07.016.

[89]  S. Hudson, M.S. Roth, T.J. Madden, R. Hudson, The effects of social media on emotions, brand relationship quality, and word of mouth: An empirical study of music festival attendees, Tourism Management. 47 (2015) 68–76. doi:10.1016/j.tourman.2014.09.001.

[90]  I. Llodrà-Riera, M.P. Martínez-Ruiz, A.I. Jiménez-Zarco, A. Izquierdo-Yusta, A multidimensional analysis of the information sources construct and its relevance for destination image formation, Tourism Management. 48 (2015) 319–328. doi:10.1016/j.tourman.2014.11.012.

[91]  G. Chareyron, J. Da-Rugna, T. Raimbault, Big data: A new challenge for tourism, in: Proceedings - 2014 IEEE International Conference on Big Data, IEEE Big Data 2014, 2014: pp. 5–7. doi:10.1109/BigData.2014.7004475.

[92]  H. Song, S.F. Witt, Forecasting international tourist flows to Macau, Tourism Management. 27 (2006) 214–224. doi:10.1016/j.tourman.2004.09.004.